\documentclass[useAMS,usenatbib]{mn2e}

\title[Electron impact excitation of Al X]{Energy levels, radiative rates and electron impact excitation rates for transitions in Al X\thanks{Tables 2 and 5 are available only in the electronic version.}}
\author[K. M. Aggarwal and F. P. Keenan]{Kanti  M.  ~Aggarwal$^{1}$\thanks{E-mail:
 K.Aggarwal@qub.ac.uk(KMA); F.Keenan@qub.ac.uk (FPK)} and Francis  P.   ~Keenan$^{1}$ \\
$^{1}$Astrophysics Research Centre, School of Mathematics and Physics, Queen's University Belfast, Belfast BT7 1NN, Northern Ireland, UK} 
\begin{document}

\date{Accepted 2013 November 21. Received 2013 November 14; in original form 2013 August 1}

\pagerange{\pageref{firstpage}--\pageref{lastpage}} \pubyear{2002}

\maketitle

\label{firstpage}

\begin{abstract}
Energy levels, radiative rates and lifetimes are calculated among the lowest 98 levels of the $n \le$ 4 configurations of Be-like Al X. The {\sc grasp} (General-purpose Relativistic Atomic Structure Package) is adopted and data are provided  for all E1, E2, M1 and M2 transitions. Similar data are also obtained with the Flexible Atomic Code ({\sc fac})  to assess the accuracy of the calculations. Based on comparisons between calculations with the two codes as well as with  available measurements,  our listed energy levels are  assessed to be accurate to better than 0.3\%. However, the accuracy for radiative rates and lifetimes  is estimated to be about  20\%. Collision strengths are also calculated for which the Dirac Atomic R-matrix Code ({\sc darc}) is used.  A wide energy range (up to 380 Ryd) is considered  and resonances  resolved in a fine energy mesh in the thresholds region. The  collision strengths are subsequently averaged over a Maxwellian velocity distribution to determine effective collision strengths up to a  temperature of 1.6$\times$10$^7$ K. Our results are compared with the previous (limited) atomic data and significant differences (up to a factor of 4) are noted for several transitions, particularly those which are not allowed in  $jj$ coupling.

\end{abstract}


\section{Introduction}
Emission lines of Al ions are widely detected in a variety of plasmas, including solar and lasing plasmas -- see for example, \cite{el1} and  \cite*{gu1}. Their observation provides diagnostics for which atomic data are required for a range of parameters, such as energy levels, radiative rates (A- values), and excitation rates or equivalently the effective collision strengths ($\Upsilon$), which are obtained from the electron impact collision strengths ($\Omega$). Emission lines of many Al ions are included in the CHIANTI database at {\tt http://www.chiantidatabase.org/}, and for Al X are listed in the 35-60,600 ${\rm \AA}$ wavelength range in the {\em Atomic Line List} (v2.04) of Peter van Hoof at ${\tt {\verb+http://www.pa.uky.edu/~peter/atomic/+}}$. Of particular interest are  two transitions, namely 2s$^2$ $^1$S$_0$--2s2p $^3$P$^o_1$ and 2s2p $^3$P$^o_1$--2p$^2$ $^1$D$_2$, at wavelengths of $\sim$ 637 and 670 $\rm \AA$, respectively.  These  have been measured in  solar spectra from the Solar Ultraviolet Measurement of Emitted Radiation (SUMER) instrument  on board  the Solar and Heliospheric Observatory (SOHO). The ratio of their intensities  is temperature sensitive (Landi et al.  2001) and hence provides an excellent diagnostic. Since these lines are  close in wavelength, they are readily detected by a single spectrograph, thus providing an additional advantage of accuracy. 
 
Unfortunately, existing atomic data for Al X  are very limited, particularly for collision strengths ($\Omega$) and effective collision strengths ($\Upsilon$).  \cite{zs92} calculated $\Omega$  for 45 transitions among the lowest 10 levels of Be-like ions with 8 $\le$ Z $\le$ 92, but did not report results for Al X. The only data available for $\Upsilon$ are those of  \cite{fpk1}, who provided analytical expressions to derive interpolated values of $\Upsilon$  based on  $R$-matrix calculations for Be-like ions between C III and Si XI. Their results are only for transitions among the lowest 10 levels of the $n$=2 configurations, and are incorporated in the CHIANTI database.  No calculation has to our knowledge  been performed with the $R$-matrix code which explicitly includes the contribution of resonances to the determination of  $\Upsilon$. This may be highly significant, particularly for the forbidden transitions, as noted earlier for another Be-like ion, i.e. Ti XIX  \citep*{tixix}.  

To analyse solar observations,  \cite{el1} adopted the atomic data for Al X  in  CHIANTI. Since these data are limited to  transitions within the $n$=2 configurations, they also calculated $\Upsilon$ for a larger model, i.e. 98 levels of the $n \le$ 4 configurations. For this  they adopted the Hebrew University Lawrence Livermore Atomic Code (HULLAC) of  \cite*{hullac},  based on the well-known and widely-used {\em distorted-wave} (DW) method. Since resonance contributions are  not normally included in DW calculations, the  results for $\Upsilon$ may be significantly underestimated. Therefore, they included their contribution using the isolated resonance approximation. As a consequence, temperatures deduced from solar observations using two different sets of atomic data are significantly different, i.e. 10$^{6.47}$ and 10$^{5.75}$ K from CHIANTI and HULLAC, respectively. 

The isolated resonance approximation takes into account the  resonance contribution  to a large extent, but   $\Upsilon$ can still be greatly underestimated, as demonstrated by  \cite{mo34,fe25} for transitions in Mo XXXIV and Fe XXV, respectively. Furthermore,   \cite{el1} did not list any atomic data and  these are not available on any website. Therefore,  in this work we report atomic data for energy levels, A- values, $\Omega$ and $\Upsilon$ for all transitions among the lowest 98 levels of the $n \le$ 4 configurations of Al X. To determine the atomic structure (i.e.  calculate energy levels and A- values) we employ the fully relativistic {\sc grasp} (General-purpose Relativistic Atomic Structure  Package) code. Our version  was  originally  developed by  \cite{grasp0} and  is referred to as GRASP0, but has been significantly revised by one of its authors (Dr. P. H. Norrington),   and is available at the website {\tt http://web.am.qub.ac.uk/DARC/}. It is a fully relativistic code,  based on the $jj$ coupling scheme, and includes higher-order relativistic corrections arising from the Breit (magnetic) interaction and quantum electrodynamics (QED) effects (vacuum polarisation and Lamb shift). Furthermore, as in our earlier work, we have used the option of {\em extended average level} (EAL). Under this option  a weighted (proportional to 2$j$+1) trace of the Hamiltonian matrix is minimised. However, the results obtained for energy levels as well as radiative rates are comparable to other options, such as {\em average level} (AL), as noted  by  \cite*{kr35,xe54}  for Kr and Xe ions. 

For the scattering calculations, we have adopted the {\em Dirac Atomic $R$-matrix Code} ({\sc darc}) of P. H. Norrington and I. P. Grant ({\tt http://web.am.qub.ac.uk/DARC/}). This is a relativistic version of the standard $R$-matrix code and is based on the $jj$ coupling scheme. For this reason, the accuracy of calculated data (for $\Omega$ and subsequently $\Upsilon$) should be higher, particularly for transitions among the {\em fine-structure} levels of a state, because resonances through the energies of degenerating levels are also taken into  account.  

\section[]{Energy levels}

The  lowest 98 levels of Al X belong to the 17  configurations: (1s$^2$) 2$\ell$2$\ell'$, 2$\ell$3$\ell'$ and 2$\ell$4$\ell'$.  Our level energies calculated from {\sc grasp},  {\em without} and {\em with} the inclusion of Breit and QED effects, are listed in Table 1.   \cite{mz} have compiled and critically evaluated experimentally-measured  energy levels of Al X, which are  available at the  NIST (National Institute of Standards and Technology)   website {\tt http://www.nist.gov/pml/data/asd.cfm}, and  included in Table 1 for comparison.   However, NIST energies are not available for many levels, particularly  of the 2$\ell$4$\ell'$ configurations, and for some of the levels their results are the same (i.e. non degenerate)   -- see for example: the 2s4f $^3$F$^o_{2,3,4}$ levels.  Also included in the table, for comparison purpose,  are energies obtained from the {\em Flexible Atomic Code} ({\sc fac}) code of \cite{fac}. These results listed under FAC1 include  the same CI (configuration interaction) as in {\sc grasp}.  

The inclusion of the Breit and QED effects (GRASP2) does not significantly alter the energies obtained with their exclusion (GRASP1), as both sets of results agree within 0.01 Ryd, and their orderings are also the same. Similarly, there is no discrepancy in the ordering of levels between GRASP and NIST, and the energy differences for common levels are generally within 0.01 Ryd. However, particularly for two levels, namely  2p3s $^1$P$^o_1$ and  2p3d $^1$F$^o_3$, the NIST energies are higher by up to 0.06 Ryd. For  some of the levels, including  2p3s $^1$P$^o_1$, the NIST energies are not highly accurate as they have placed a question mark on these.  

 Our FAC1 energies  show the same ordering as those of  GRASP and NIST, and agree with our GRASP2 calculations within 0.05 Ryd.  The inclusion of the 2$\ell$5$\ell'$ configurations, labelled FAC2 calculations in Table 1, makes no appreciable difference. Small discrepancies in the {\sc grasp} and {\sc fac} energies, also noted for several other ions,  are primarily due to the different ways that the  calculations of central potential for radial orbitals and recoupling schemes of angular parts have been performed -- see the detailed discussion in the {\sc fac} manual ({\tt {\verb+http://sprg.ssl.berkeley.edu/~mfgu/fac/+}}). However,  for the levels in  common our GRASP2 energies are slightly closer to those of NIST than  those from  FAC. Therefore, based on the comparisons shown,  our GRASP2 energy levels listed in Table 1 are assessed  to be accurate to better than 0.3\%.

\section{Radiative rates}

Since the  A- values of  \cite{zs92} are limited to E1 transitions among the lowest 10 levels of  Al X, we here provide a complete set of data for all transitions among the 98 levels. Furthermore, A- values are calculated for four types of transitions, namely electric dipole (E1), electric quadrupole (E2), magnetic dipole (M1), and  magnetic quadrupole (M2). Generally, E1 transitions are dominant, but occasionally other types of transitions are also prominent and are therefore  required for  a  complete plasma model. The absorption oscillator strength ($f_{ij}$) and radiative rate A$_{ji}$ (in s$^{-1}$) for all types of  transition $i \to j$ are related by the following expression  \citep*{rhg}:

\begin{equation}
f_{ij} = \frac{mc}{8{\pi}^2{e^2}}{\lambda^2_{ji}} \frac{{\omega}_j}{{\omega}_i} A_{ji}
 = 1.49 \times 10^{-16} \lambda^2_{ji} (\omega_j/\omega_i) A_{ji}
\end{equation}
where $m$ and $e$ are the electron mass and charge, respectively, $c$ the velocity of light,  $\lambda_{ji}$  the transition energy/wavelength in $\rm \AA$, and $\omega_i$ and $\omega_j$  the statistical weights of the lower ($i$) and upper ($j$) levels, respectively. However, the relationships between oscillator strength f$_{ij}$ (dimensionless) and the line strength S (in atomic unit, 1 a.u. = 6.460$\times$10$^{-36}$ cm$^2$ esu$^2$) with the A- values are different for different types of transitions -- see Eqs. (2--5) of  \cite{tixix}.

In Table 2 we list transition energies/wavelengths ($\lambda$, in $\rm \AA$), radiative rates (A$_{ji}$, in s$^{-1}$), oscillator strengths (f$_{ij}$, dimensionless), and line strengths (S, in a.u.) for all 1468 electric dipole (E1) transitions among the 98 levels of  Al X.  For conciseness,  results are only listed in the length  form, which are considered to be more accurate. However, below we discuss  the velocity/length  form ratio, as this provides some assessment of  the accuracy of the results. Also note that the {\em indices} used  to represent the lower and upper levels of a transition are defined in Table 1. There  are 1754 electric quadrupole (E2), 1424  magnetic dipole (M1), and 1792 magnetic quadrupole (M2) transitions  among the same 98 levels. However, for these only the A-values are listed in Table 2, and the corresponding results for f-  values can be easily obtained through Eq. (1).

As noted in section 1,  f-values  for  Al X  are  available in the literature \citep*{zs92} for only a limited number of  transitions.  Therefore, as with energy levels we have also performed a calculation with the {\sc fac} code of \cite{fac}.  Our calculated  f- values from both  {\sc grasp} and {\sc fac}, as well as their ratio,  are listed in Table 3  for some representative transitions among the lowest 20 levels of  Al X.  For these transitions the agreement between the two sets of f- values is better than 20\%, particularly for the  strong ones with  f $\ge$ 0.01. However,  there are  4 exceptions, namely 21--71 (2p3s $^3$P$^o_0$ -- 2p4p $^3$P$_1$), 31--85 (2p3p $^3$P$_1$ -- 2p4d $^3$P$^o_2$),  46--76 (2p3d $^1$P$^o_1$ -- 2p4p $^1$D$_2$), and 46--82 (2p3d $^1$P$^o_1$ -- 2p4f $^3$F$_2$), for which the discrepancies are larger, but still less than 50\%.  These discrepancies are  partly due to the corresponding differences in the energy levels of the two calculations. 

A general criterion to assess the accuracy of  f- or A- values is to compare the ratio of their length and velocity forms. This should ideally be close to unity but often is not,  because the two formulations are not exactly the same.  Therefore, we also  include in Table 3  the ratio of the velocity and length forms.  For a majority (89\%) of the strong E1 transitions (f $\ge$ 0.01) the ratio is within 20\% of  unity, but discrepancies for  some are higher,  although mostly within a factor of two.  However, for a few ($\sim$ 8\%) weak(er) transitions (f $\le$ 10$^{-3}$) the two forms of the f- value differ by up to several  orders of magnitude -- see for  example: 34--37, 56--58, 67-75, and 71--73. For all of these transitions, $\Delta$E$_{ij}$ is very low and a small variation in this has a large effect on the  f- value.  A few other transitions with significant $\Delta$E$_{ij}$ for which the two forms disagree by over 20\% are: 3--12 (f $\sim$ 10$^{-6}$), 6--15 (f $\sim$ 10$^{-3}$), and 10--15 (f $\sim$ 10$^{-5}$), as shown in Table 3.  Finally, as noted for the energy levels in section 2, inclusion of  additional CI with the  $n$ = 5 configurations does not make any appreciable effect on the f- or A- values. To be specific, discrepancies in A- values for all strong E1 transitions  are less than $\sim$ 20\% with those listed in Tables 2 and 3. Therefore, based on this and other comparisons already discussed,  we are confident that for almost all strong E1 transitions listed in Table 2, our f- values (and other related parameters)  are accurate to better than 20\%. However, for the weaker E1 and other types of  transitions (i.e. E2, M1 and M2) the accuracy may be comparatively lower. Finally, we note that these conclusions are similar to those arrived earlier for transitions of another Be-like ion, i.e. Ti XIX \citep*{tixix}.

\section{Lifetimes}

The lifetime $\tau$ for a level $j$ is defined as follows (Woodgate  1970):

\begin{equation}  {\tau}_j = \frac{1}{{\sum_{i}^{}} A_{ji}}.  
\end{equation} 
Since this is a measurable quantity, it facilitates an assessment of  the accuracy of the A- values. Therefore, in Table 1 we have also listed our calculated lifetimes. Generally, A- values for E1 transitions  dominate, but for higher accuracy we have also  included the contributions from E2, M1 and M2.  Their  inclusion is particularly useful for those levels which do not connect via E1 transitions. \cite{et1} have measured the lifetime of the 2s2p $^1$P$^o_1$ level to be 175$\pm$15 ps, which compares very well with our result of 160 ps and the 173 ps theoretical value of Andersson et al. (2009). They have also measured lifetimes corresponding to the  2s2p $^1$P$^o_1$--2p$^2$ $^1$D$_2$ (5--9) and 2s2p $^1$P$^o_1$--2p$^2$ $^1$S$_0$ (5--10) transitions to be 1080$\pm$80 and 112$\pm$12 ps, respectively, which compare favourably  with our results of 1072 and 105 ps, respectively.
 
 \section{Collision strengths}

The collision strength for electron impact excitation  ($\Omega$) is related to the better-known parameter collision cross section ($\sigma_{ij}$, $\pi{a_0}^2$) by the following equation  \citep*{bt}:

\begin{equation}
\Omega_{ij}(E) = {k^2_i}\omega_i\sigma_{ij}(E)
\end{equation}
where ${k^2_i}$ is the incident energy of the electron and $\omega_i$ is the statistical weight of the initial state.  Since $\Omega$  is a symmetric and dimensionless quantity, results for it are preferred over those of $\sigma_{ij}$.

As in our earlier work, such as on Ti XIX \citep*{tixix}, for calculating  $\Omega$ we have adopted the {\em Dirac Atomic $R$-matrix Code} ({\sc darc}) of P. H. Norrington and I. P. Grant,  available  at the website {\tt http://web.am.qub.ac.uk/DARC/}.  This code includes the relativistic effects, which are very important for high Z ions, but are equally significant for taking into account degeneracy among the fine-structure levels of a state of an ion with lower Z, such as Al X.  The {\sc darc} code is based on the $jj$ coupling scheme and uses the  Dirac-Coulomb Hamiltonian in an $R$-matrix approach. The $R$-matrix radius adopted for  Al X is 6.4 au, and 55  continuum orbitals have been included for each channel angular momentum in the expansion of the wavefunction. This large expansion is computationally more demanding but allows us to compute $\Omega$ up to an energy of  380 Ryd,  $\sim$355 Ryd {\em above} the highest threshold considered in this work. Furthermore, this large energy range is  sufficient to calculate values of effective collision strength $\Upsilon$ (see section 6)  up to T$_e$ = 1.8 $\times$10$^{7}$ K, more than an order of magnitude higher than  the temperature of maximum abundance in ionisation equilibrium for  Al X, i.e. 1.3 $\times$10$^{6}$ K  \citep*{pb}.  The maximum number of channels for a partial wave is 428 and the corresponding size of the Hamiltonian matrix is 23,579. To achieve  convergence of  $\Omega$ for a majority of transitions and at all energies, we have included all partial waves with angular momentum $J \le$ 40.5.  Additionally, to account for higher neglected partial waves, we have included a top-up, based on the Coulomb-Bethe  \citep*{ab} and  geometric series  approximations for allowed and forbidden transitions, respectively. These contributions enhance the accuracy of our calculated values of $\Omega$s, particularly at the higher end of the energy range.

In Table 4 we list our values of $\Omega$ for resonance transitions of  Al X at energies {\em above} thresholds. The  indices used  to represent the levels of a transition have already been defined in Table 1. Unfortunately, no similar data are available for comparison purposes as already noted in section 1. One way to assess the accuracy of these results is to compare with the similar calculations from {\sc fac}, as undertaken in our work on Ti XIX \citep*{tixix}. However, such a comparison  is not very useful, particularly for the forbidden transitions,  because  often there are anomalies in the FAC calculations, as shown in  Fig. 6 of  \cite{mgxi,caxix}. Nevertheless, our listed results for $\Omega$ should be helpful for future comparisons. 

\section{Effective collision strengths}

As well as  energy levels and radiative rates, excitation and de-excitation rates are required for plasma modelling, which are determined from the collision strengths ($\Omega$). However, $\Omega$ does not vary smoothly with increasing energy, particularly at energies in between  the thresholds. The  threshold energy region  is  generally dominated by numerous closed-channel (Feshbach) resonances, especially for (semi) forbidden transitions. Therefore, values of $\Omega$ should be calculated in a fine energy mesh  to accurately account for their contribution. Additionally, in many plasmas electrons have a wide distribution of velocities, and therefore it is more appropriate to average values of $\Omega$  over a suitable distribution. For astrophysical applications the most appropriate and commonly used distribution is {\em Maxwellian}, although any other distribution may also be applied if suitable for a particular plasma.  Such an averaged value, known as {\em effective} collision strength ($\Upsilon$)  \citep*{bt} is:

\begin{equation}
\Upsilon(T_e) = \int_{0}^{\infty} {\Omega}(E) \, {\rm exp}(-E_j/kT_e) \,d(E_j/{kT_e}),
\end{equation}
where $k$ is Boltzmann constant, T$_e$  electron temperature in K, and E$_j$  the electron energy with respect to the final (excited) state. Once the value of $\Upsilon$ is
known the corresponding results for the excitation q(i,j) and de-excitation q(j,i) rates can be easily obtained from the following equations:

\begin{equation}
q(i,j) = \frac{8.63 \times 10^{-6}}{{\omega_i}{T_e^{1/2}}} \Upsilon \, {\rm exp}(-E_{ij}/{kT_e}) \hspace*{1.0 cm}{\rm cm^3s^{-1}}
\end{equation}
and
\begin{equation}
q(j,i) = \frac{8.63 \times 10^{-6}}{{\omega_j}{T_e^{1/2}}} \Upsilon \hspace*{1.0 cm}{\rm cm^3 s^{-1}},
\end{equation}
where $\omega_i$ and $\omega_j$ are the statistical weights of the initial ($i$) and final ($j$) states, respectively, and E$_{ij}$ is the transition energy. The contribution of resonances often enhances the values of $\Upsilon$ over those of the background  collision strengths ($\Omega_B$), particularly for the (semi) forbidden transitions. This enhancement can be  dominant (by up to an order of magnitude or even more), but  depends on the type of a transition as well as the temperature.  Generally, the enhancement in $\Upsilon$ is greater at lower temperatures. Similarly, values of $\Omega$ should  be calculated over a wide energy range (above thresholds)  to obtain convergence of the integral in Eq. (4), as demonstrated in Fig. 7 of  \cite{ni11}. If calculations of $\Omega$ are performed only for a limited range of energy, it is still necessary to include the contribution of  $\Omega$  at  high energies. For this  the high energy limits recommended by  \cite{bt} for a range of transitions may be adopted. However, in our work  there is no such need  because calculations for $\Omega$ have already  been performed up to sufficiently high energies, as clarified in section 5. 

To delineate resonances, we have performed our calculations of $\Omega$ at over $\sim$ 13,000 energies in the thresholds region. Close to thresholds ($\sim$0.1 Ryd above a threshold) the energy mesh is 0.001 Ryd, and away from thresholds is 0.002 Ryd.  This fine resolution accounts for the majority of resonances, and their density and importance can be appreciated from transitions of Ti XIX shown in Figs. 7--12 of  \cite{tixix}.  For transitions of Al X  we observe similar dense resonances, as shown  in Figs. 1 and 2  for two important lines, namely 2s$^2$ $^1$S$_0$--2s2p $^3$P$^o_1$ (1--3) and 2s2p $^3$P$^o_1$--2p$^2$ $^1$D$_2$ (3--9). Both of these are allowed transitions in $jj$ coupling and yet resonances  are not only dense but also highly significant in  magnitude. Furthermore, resonances   are spread over a range of 20 Ryd (equivalently over 3$\times$10$^6$ K), and hence  make an appreciable  contribution to  $\Upsilon$  over the entire range of  temperatures of interest for transitions of Al X. 

Our calculated values of $\Upsilon$ are listed in Table 5 over a wide temperature range up to 10$^{7.2}$ K, suitable for applications to a wide range  of laboratory and astrophysical plasmas. Corresponding data at any desired temperature can either be easily interpolated, because $\Upsilon$ is a slowly varying function of T$_e$,  or may be requested from  the first author. As noted in section 1,  \cite{fpk1}  have reported values of $\Upsilon$ for transitions among the lowest 10 levels of Al X. In Table 6 we compare  results for $\Upsilon$  at three  temperatures of 10$^{5.9}$, 10$^{6.1}$ and 10$^{6.3}$ K, which are most relevant for Al X  \citep{pb}. The interpolated values of $\Upsilon$  listed by Keenan et al. are based on the $R$-matrix calculations for C III, O V, Ne VII and Si XI, which were performed in $LS$ coupling. Since their work is not based on direct calculations for  Al X, differences with their results are not unexpected. For  transitions which are allowed in  $LS$ coupling, such as 1--5 and  5--9/10, there is no discrepancy between our results and those of  Keenan et al. This is because such allowed transitions do not normally have a significant contribution from resonances. However, there are differences between the two sets of $\Upsilon$ of  up a factor of two for several (but not all) transitions which are allowed in $LS$ coupling but not in $jj$, such as  2/3/4--6/7/8, i.e. 2s2p $^3$P$^o_{0,1,2}$ -- 2p$^2$ $^3$P$_{0,1,2}$. For other transitions (particularly forbidden) the $\Upsilon$ values of Keenan et al.  are {\em underestimated} by up to a factor of 4.

\section{Conclusions}

Energies and lifetimes for the lowest 98 levels of  Al X belonging to the $n \le$ 4 configurations have been reported, for which the {\sc grasp} code has been adopted. Also listed are results for radiative rates for four types of transitions (E1, E2, M1 and M2). Based on a variety of comparisons with available measurements, as well as with analogous calculations with the  {\sc fac} code, our results for radiative rates, oscillator strengths and line strengths are judged to be accurate to better than 20\% for a majority of  strong transitions. Similarly, energy levels are assessed to be accurate to $\sim$0.3\%. Measurements of lifetimes are available for only three levels for which there are no discrepancies with theory.

Results have also been reported for collision strengths  over a wide range of energy, but only for resonance transitions. However, corresponding results for effective collision strengths are listed for {\em all} transitions among the 98 levels of Al X and over a wide range of temperature, suitable for applications in a variety of plasmas. For  calculations of $\Upsilon$, resonances in the thresholds energy region for many transitions are noted to be as dominant as for Ti XIX  \citep*{tixix}. Their inclusion in the determination of $\Upsilon$  has significantly enhanced the results. Since no prior  calculations with comparable accuracy and complexity are available, it is  not straightforward to assess the uncertainty of our values of $\Upsilon$.  However, comparisons between our results  and those interpolated by   \cite{fpk1} have been made for transitions among the lowest 10 levels, and the latter are found to be underestimated by up to a factor of four for several transitions, mostly forbidden. 

Since a large range of partial waves  has been considered to achieve convergence of $\Omega$ at all energies  and contribution of higher neglected partial waves has been included, our results for $\Omega$ should be accurate to better than 20\%. This assessment of accuracy is mainly based on comparisons of similar data for  Be-like ion of Ti. Similarly,  to  calculate  values of $\Upsilon$ up to T$_e$ = 10$^{7.2}$ K,  we have  included a wide energy range for $\Omega$ and have also  resolved resonances in a fine energy mesh to account for their contribution. Hence, we see no  apparent limitations in our  data. Moreover, as for  Ti XIX   \citep*{tixix}, we estimate the accuracy of our results for  $\Upsilon$  to be better than 20\% for most transitions.  However, there is scope for improvement, especially  for transitions involving  levels of the $n$ = 4 configurations. This can perhaps be achieved  by the inclusion of  levels of the $n$ = 5 configurations  in the collisional calculations.  At present we believe the reported results  for radiative and excitation rates for transitions in  Al X are the most exhaustive and accurate available to date. The complete set of atomic data should be highly useful for modelling astrophysical and fusion plasmas. 

\section*{Acknowledgments}

KMA is thankful to  AWE Aldermaston for    financial support.

\clearpage
\newpage

\setcounter{table}{0}                                                                                         
\begin{table*}                                                                                                
\caption{ Energy levels (in Ryd) of Al X and their lifetimes.} 

\begin{tabular}{rllrrrrrl} \hline \hline
Index  & Configuration       & Level          &  NIST      &   GRASP1    & GRASP2   &  FAC1     & FAC2     & $\tau$ (s) \\
 \hline
   1 &  2s$^2$    &  $^1$S$  _0$  &   0.00000	  &   0.00000    &   0.00000 	&   0.00000  &   0.00000   &   ........    \\
   2 &  2s2p	  &  $^3$P$^o_0$  &   1.41381	  &   1.41694    &   1.41897 	&   1.42722  &   1.42661   &   ........    \\ 
   3 &  2s2p	  &  $^3$P$^o_1$  &   1.42885	  &   1.43399    &   1.43387 	&   1.44188  &   1.44128   &   5.963-06    \\ 
   4 &  2s2p	  &  $^3$P$^o_2$  &   1.46194	  &   1.46944    &   1.46662 	&   1.47416  &   1.47357   &   1.246-00    \\ 
   5 &  2s2p	  &  $^1$P$^o_1$  &   2.73827	  &   2.81662    &   2.81633 	&   2.81532  &   2.81053   &   1.600-10    \\ 
   6 &  2p$^2$    &  $^3$P$  _0$  &   3.68675	  &   3.71751    &   3.71907 	&   3.73711  &   3.73633   &   2.154-10    \\ 
   7 &  2p$^2$    &  $^3$P$  _1$  &   3.70446	  &   3.73616    &   3.73652 	&   3.75430  &   3.75353   &   2.129-10    \\ 
   8 &  2p$^2$    &  $^3$P$  _2$  &   3.73337	  &   3.76904    &   3.76518 	&   3.78247  &   3.78171   &   2.096-10    \\
   9 &  2p$^2$    &  $^1$D$  _2$  &   4.09826	  &   4.17709    &   4.17504 	&   4.19138  &   4.18716   &   1.050-09    \\ 
  10 &  2p$^2$    &  $^1$S$  _0$  &   5.04644	  &   5.17695    &   5.17938 	&   5.18781  &   5.18508   &   1.052-10    \\ 
  11 &  2s3s	  &  $^3$S$  _1$  &   16.9109	  &  16.90263    &  16.89542 	&  16.89461  &  16.89432   &   5.201-12    \\
  12 &  2s3s	  &  $^1$S$  _0$  &   17.1721	  &  17.16367    &  17.15693 	&  17.17011  &  17.16978   &   1.435-11    \\
  13 &  2s3p	  &  $^1$P$^o_1$  &   17.5314	  &  17.53546    &  17.52796 	&  17.53879  &  17.53775   &   2.139-12    \\
  14 &  2s3p	  &  $^3$P$^o_0$  &  	     	  &  17.55638    &  17.55008 	&  17.56230  &  17.56240   &   4.322-10    \\
  15 &  2s3p	  &  $^3$P$^o_1$  &   	     	  &  17.56265    &  17.55555 	&  17.56749  &  17.56755   &   4.143-11    \\
  16 &  2s3p	  &  $^3$P$^o_2$  &  	     	  &  17.57084    &  17.56323 	&  17.57501  &  17.57512   &   4.037-10    \\
  17 &  2s3d	  &  $^3$D$  _1$  &   17.9142	  &  17.91422    &  17.90599 	&  17.92123  &  17.91911   &   1.076-12    \\  
  18 &  2s3d	  &  $^3$D$  _2$  &   17.9162	  &  17.91616    &  17.90756 	&  17.92272  &  17.92059   &   1.079-12    \\
  19 &  2s3d	  &  $^3$D$  _3$  &   17.9182	  &  17.91907    &  17.91022 	&  17.92529  &  17.92316   &   1.084-12    \\
  20 &  2s3d	  &  $^1$D$  _2$  &   18.1555	  &  18.18722    &  18.17882 	&  18.19304  &  18.18938   &   1.608-12    \\
  21 &  2p3s	  &  $^3$P$^o_0$  &  	     	  &  18.69695    &  18.69110 	&  18.71734  &  18.71745   &   6.603-12    \\
  22 &  2p3s	  &  $^3$P$^o_1$  &  	     	  &  18.71325    &  18.70615 	&  18.73218  &  18.73226   &   6.538-12    \\
  23 &  2p3s	  &  $^3$P$^o_2$  &   18.7460	  &  18.75178    &  18.74171 	&  18.76709  &  18.76720   &   6.422-12    \\
  24 &  2p3s	  &  $^1$P$^o_1$  &   19.0625	  &  19.01237    &  19.00401 	&  19.03602  &  19.03186   &   5.332-12    \\
  25 &  2p3p	  &  $^1$P$  _1$  &   19.0894	  &  19.09350    &  19.08552 	&  19.11202  &  19.11216   &   3.777-12    \\
  26 &  2p3p	  &  $^3$D$  _1$  &   19.1578	  &  19.16750    &  19.16008 	&  19.18591  &  19.18569   &   7.090-12    \\
  27 &  2p3p	  &  $^3$D$  _2$  &   19.1721	  &  19.18332    &  19.17515 	&  19.20029  &  19.20005   &   7.590-12    \\
  28 &  2p3p	  &  $^3$D$  _3$  &   19.0240	  &  19.21937    &  19.20844 	&  19.23319  &  19.23295   &   7.516-12    \\
  29 &  2p3p	  &  $^3$S$  _1$  &   19.3160	  &  19.33016    &  19.32141 	&  19.34755  &  19.34700   &   4.454-12    \\
  30 &  2p3p	  &  $^3$P$  _0$  &   	     	  &  19.39104    &  19.38473 	&  19.43306  &  19.43188   &   4.346-12    \\
  31 &  2p3p	  &  $^3$P$  _1$  &   19.3980	  &  19.40812    &  19.40025 	&  19.44680  &  19.44568   &   4.348-12    \\
  32 &  2p3p	  &  $^3$P$  _2$  &   19.4137	  &  19.42677    &  19.41709 	&  19.46398  &  19.46286   &   4.347-12    \\
  33 &  2p3d	  &  $^3$F$^o_2$  &  	     	  &  19.47683    &  19.46968 	&  19.50477  &  19.50257   &   1.062-11    \\
  34 &  2p3d	  &  $^3$F$^o_3$  &  	     	  &  19.50513    &  19.49594 	&  19.53497  &  19.53229   &   1.144-10    \\
  35 &  2p3d	  &  $^1$D$^o_2$  &   19.5155	  &  19.52021    &  19.50998 	&  19.54890  &  19.54802   &   3.121-12    \\
  36 &  2p3d	  &  $^3$F$^o_4$  &  	     	  &  19.53388    &  19.52225 	&  19.55770  &  19.55499   &   2.471-09    \\
  37 &  2p3p	  &  $^1$D$  _2$  &   19.5778	  &  19.61598    &  19.60698 	&  19.66078  &  19.65660   &   2.935-12    \\
  38 &  2p3d	  &  $^3$D$^o_1$  &   19.6893	  &  19.69761    &  19.68861 	&  19.72802  &  19.72823   &   8.663-13    \\
  39 &  2p3d	  &  $^3$D$^o_2$  &   19.7012	  &  19.70542    &  19.69591 	&  19.73859  &  19.73878   &   8.831-13    \\
  40 &  2p3d	  &  $^3$D$^o_3$  &   19.7138	  &  19.72105    &  19.70995 	&  19.75056  &  19.75077   &   8.622-13    \\
  41 &  2p3d	  &  $^3$P$^o_2$  &   19.7762	  &  19.79012    &  19.77978 	&  19.81691  &  19.81661   &   1.553-12    \\
  42 &  2p3d	  &  $^3$P$^o_1$  &   19.7898	  &  19.80156    &  19.79099 	&  19.82829  &  19.82796   &   1.579-12    \\
  43 &  2p3d	  &  $^3$P$^o_0$  &  	     	  &  19.80750    &  19.79743 	&  19.83558  &  19.83523   &   1.606-12    \\
  44 &  2p3p	  &  $^1$S$  _0$  &  	     	  &  19.94986    &  19.94280 	&  20.00059  &  19.98775   &   5.527-12    \\
  45 &  2p3d	  &  $^1$F$^o_3$  &   19.9828	  &  20.03858    &  20.02755 	&  20.06982  &  20.06305   &   7.032-13    \\
  46 &  2p3d	  &  $^1$P$^o_1$  &   	     	  &  20.08601    &  20.07597 	&  20.11448  &  20.11208   &   1.174-12    \\
  47 &  2s4s	  &  $^3$S$  _1$  &  	     	  &  22.54933    &  22.54097 	&  22.54646  &  22.54537   &   1.037-11    \\
  48 &  2s4s	  &  $^1$S$  _0$  &  	     	  &  22.65071    &  22.64271 	&  22.65172  &  22.64843   &   1.106-11    \\
 \hline
\end{tabular} 
\end{table*}

\clearpage
\newpage 
\setcounter{table}{0}                                                                                         
\begin{table*}                                                                                                
\caption{ Energy levels (in Ryd) of Al X and their lifetimes.} 

\begin{tabular}{rllrrrrrl} \hline \hline
Index  & Configuration       & Level          &  NIST      &   GRASP1    & GRASP2   &  FAC1     & FAC2     & $\tau$ (s) \\
 \hline
  49 &  2s4p	  &  $^3$P$^o_0$  &  	       &   22.79976    &   22.79178   &  22.80264  &  22.80264   &  3.219-11 	\\
  50 &  2s4p	  &  $^3$P$^o_1$  &  	       &   22.80151    &   22.79331   &  22.80412  &  22.80410   &  3.045-11 	\\
  51 &  2s4p	  &  $^3$P$^o_2$  &  	       &   22.80575    &   22.79723   &  22.80786  &  22.80787   &  3.258-11 	\\
  52 &  2s4p	  &  $^1$P$^o_1$  &   	       &   22.83208    &   22.82350   &  22.83676  &  22.83426   &  3.598-12 	\\
  53 &  2s4d	  &  $^3$D$  _1$  &   	       &   22.94439    &   22.93591   &  22.94425  &  22.94326   &  2.738-12 	\\
  54 &  2s4d	  &  $^3$D$  _2$  &   	       &   22.94507    &   22.93645   &  22.94479  &  22.94380   &  2.743-12 	\\
  55 &  2s4d	  &  $^3$D$  _3$  &   	       &   22.94610    &   22.93741   &  22.94574  &  22.94474   &  2.750-12 	\\
  56 &  2s4d	  &  $^1$D$  _2$  &  23.0328   &   23.03637    &   23.02777   &  23.03406  &  23.03223	 &  3.107-12    \\
  57 &  2s4f	  &  $^3$F$^o_2$  &  23.0420   &   23.03790    &   23.02937   &  23.03614  &  23.03421   &  6.766-12 	\\
  58 &  2s4f	  &  $^3$F$^o_3$  &  23.0420   &   23.03825    &   23.02961   &  23.03639  &  23.03446   &  6.766-12 	\\
  59 &  2s4f	  &  $^3$F$^o_4$  &  23.0420   &   23.03872    &   23.03003   &  23.03679  &  23.03485   &  6.767-12 	\\
  60 &  2s4f	  &  $^1$F$^o_3$  &  	       &   23.06257    &   23.05392   &  23.06192  &  23.05970   &  6.749-12    \\
  61 &  2p4s	  &  $^3$P$^o_0$  &  	       &   24.21374    &   24.20663   &  24.23547  &  24.23546   &  1.115-11 	\\
  62 &  2p4s	  &  $^3$P$^o_1$  &  	       &   24.22541    &   24.21768   &  24.24657  &  24.24586   &  1.071-11 	\\
  63 &  2p4s	  &  $^3$P$^o_2$  &  	       &   24.26942    &   24.25816   &  24.28604  &  24.28604   &  1.066-11 	\\
  64 &  2p4s	  &  $^1$P$^o_1$  &  	       &   24.33218    &   24.32131   &  24.35227  &  24.34046   &  7.459-12 	\\
  65 &  2p4p	  &  $^1$P$  _1$  &  	       &   24.39368    &   24.38599   &  24.41716  &  24.41709   &  5.881-12 	\\
  66 &  2p4p	  &  $^3$D$  _1$  &  	       &   24.42626    &   24.41706   &  24.44825  &  24.44790   &  6.445-12 	\\
  67 &  2p4p	  &  $^3$D$  _2$  &  	       &   24.42918    &   24.42038   &  24.45195  &  24.45182   &  7.901-12 	\\
  68 &  2p4p	  &  $^3$D$  _3$  &   	       &   24.46580    &   24.45399   &  24.48457  &  24.48453   &  7.977-12 	\\
  69 &  2p4p	  &  $^3$S$  _1$  &  	       &   24.48873    &   24.47906   &  24.51326  &  24.50573   &  6.182-12 	\\
  70 &  2p4p	  &  $^3$P$  _0$  &  	       &   24.49044    &   24.48248   &  24.52186  &  24.52000   &  7.285-12 	\\
  71 &  2p4p	  &  $^3$P$  _1$  &  	       &   24.52093    &   24.50993   &  24.54626  &  24.54261   &  6.635-12 	\\
  72 &  2p4p	  &  $^3$P$  _2$  &  	       &   24.52528    &   24.51423   &  24.55234  &  24.55105   &  7.233-12 	\\
  73 &  2p4d	  &  $^3$F$^o_2$  &  	       &   24.53974    &   24.53225   &  24.56341  &  24.56166   &  9.250-12 	\\
  74 &  2p4d	  &  $^3$F$^o_3$  &  	       &   24.56480    &   24.55592   &  24.58731  &  24.58512   &  1.027-11 	\\
  75 &  2p4d	  &  $^1$D$^o_2$  &   	       &   24.57451    &   24.56504   &  24.59586  &  24.59471   &  5.064-12 	\\
  76 &  2p4p	  &  $^1$D$  _2$  &  24.5755   &   24.59319    &   24.58233   &  24.62476  &  24.61978   &  5.473-12 	\\
  77 &  2p4d	  &  $^3$F$^o_4$  &  	       &   24.59736    &   24.58565   &  24.61617  &  24.61389   &  1.423-11 	\\
  78 &  2p4d	  &  $^3$D$^o_1$  &  	       &   24.61927    &   24.61082   &  24.64026  &  24.63980   &  2.096-12 	\\
  79 &  2p4d	  &  $^3$D$^o_2$  &  	       &   24.63018    &   24.62033   &  24.65013  &  24.64908   &  2.414-12 	\\
  80 &  2p4f	  &  $^1$F$  _3$  &  	       &   24.63738    &   24.62916   &  24.65589  &  24.65263   &  6.803-12 	\\
  81 &  2p4f	  &  $^3$F$  _3$  &  	       &   24.64055    &   24.63240   &  24.65934  &  24.65823   &  6.938-12 	\\
  82 &  2p4f	  &  $^3$F$  _2$  &  	       &   24.64195    &   24.63346   &  24.66072  &  24.66063   &  6.880-12 	\\
  83 &  2p4f	  &  $^3$F$  _4$  &  	       &   24.64430    &   24.63607   &  24.66310  &  24.65907   &  7.045-12 	\\
  84 &  2p4d	  &  $^3$D$^o_3$  &  	       &   24.64876    &   24.63750   &  24.66663  &  24.66666   &  2.100-12 	\\
  85 &  2p4d	  &  $^3$P$^o_2$  &   	       &   24.67077    &   24.65945   &  24.68876  &  24.68505   &  2.914-12 	\\
  86 &  2p4d	  &  $^3$P$^o_1$  &  	       &   24.67783    &   24.66650   &  24.69593  &  24.69122   &  3.141-12 	\\
  87 &  2p4d	  &  $^3$P$^o_0$  &  	       &   24.68175    &   24.67065   &  24.70015  &  24.69477   &  3.344-12 	\\
  88 &  2p4f	  &  $^3$G$  _3$  &  	       &   24.68824    &   24.67701   &  24.70310  &  24.69764   &  6.913-12 	\\
  89 &  2p4f	  &  $^3$G$  _4$  &  	       &   24.69295    &   24.68171   &  24.70791  &  24.70163   &  7.024-12 	\\
  90 &  2p4f	  &  $^3$G$  _5$  &  	       &   24.70949    &   24.69767   &  24.72356  &  24.71287   &  6.980-12 	\\
  91 &  2p4f	  &  $^3$D$  _3$  &  	       &   24.71783    &   24.70694   &  24.73342  &  24.75159   &  6.831-12 	\\
  92 &  2p4f	  &  $^3$D$  _2$  &  	       &   24.72363    &   24.71272   &  24.73956  &  24.73334   &  6.818-12 	\\
  93 &  2p4f	  &  $^1$G$  _4$  &  	       &   24.72378    &   24.71209   &  24.73907  &  24.72144   &  7.701-12 	\\
  94 &  2p4f	  &  $^3$D$  _1$  &  	       &   24.73681    &   24.72532   &  24.75171  &  24.72754   &  6.820-12 	\\
  95 &  2p4p	  &  $^1$S$  _0$  &  	       &   24.74104    &   24.73058   &  24.78352  &  24.73931   &  1.064-11 	\\
  96 &  2p4f	  &  $^1$D$  _2$  &  	       &   24.74808    &   24.73671   &  24.76430  &  24.76358   &  6.805-12 	\\
  97 &  2p4d	  &  $^1$F$^o_3$  &   	       &   24.77573    &   24.76474   &  24.79181  &  24.77893   &  1.472-12 	\\
  98 &  2p4d	  &  $^1$P$^o_1$  &  	       &   24.78736    &   24.77677   &  24.80363  &  24.79443   &  2.300-12 	\\
 \hline	
\end{tabular}

\begin{flushleft}
{\small
NIST: {\tt http://www.nist.gov/pml/data/asd.cfm} \\
GRASP1: Coulomb energies \\
GRASP2: QED corrected energies \\
FAC1: Energies from the FAC for 98 level calculations\\
FAC2: Energies from the FAC for 166 level calculations \\
}
\end{flushleft}
\end{table*} 


\setcounter{table}{2}         
\clearpage
\newpage                                                                                 
\begin{table*}                                                                                                
\caption{Comparison between GRASP and FAC  f- values for some transitions of  Al X. ($a{\pm}b \equiv$ $a\times$10$^{{\pm}b}$).}            
\begin{tabular}{rrlllll}                                                                                    
\hline                                                                                                        
\hline                                                                                                                                                                                                               
  $i$ & $j$ &   f (GRASP)    &    f (FAC)    & Vel./Len. & f(GRASP)/f(FAC)    \\					       
\hline                                          					      
    1 &     3 &    3.046$-$5  &   3.039$-$5  &  0.70  &  1.00 \\
    1 &     5 &    2.942$-$1  &   2.942$-$1  &  0.97  &  1.00 \\
    1 &    13 &    5.324$-$1  &   5.445$-$1  &  0.97  &  0.98 \\
    1 &    15 &    2.478$-$2  &   2.183$-$2  &  0.97  &  1.14 \\
    2 &     7 &    1.124$-$1  &   1.127$-$1  &  0.90  &  1.00 \\
    2 &    11 &    3.321$-$2  &   3.394$-$2  &  0.94  &  0.98 \\
    2 &    17 &    7.100$-$1  &   7.095$-$1  &  0.98  &  1.00 \\
    3 &     6 &    3.688$-$2  &   3.698$-$2  &  0.90  &  1.00 \\
    3 &     7 &    2.790$-$2  &   2.797$-$2  &  0.90  &  1.00 \\
    3 &     8 &    4.719$-$2  &   4.731$-$2  &  0.91  &  1.00 \\
    3 &     9 &    2.755$-$5  &   2.640$-$5  &  0.61  &  1.04 \\
    3 &    10 &    4.082$-$6  &   4.097$-$6  &  1.70  &  1.00 \\
    3 &    11 &    3.334$-$2  &   3.404$-$2  &  0.94  &  0.98 \\
    3 &    12 &    1.076$-$6  &   1.377$-$6  &  0.46  &  0.78 \\
    3 &    17 &    1.775$-$1  &   1.773$-$1  &  0.98  &  1.00 \\
    3 &    18 &    5.318$-$1  &   5.315$-$1  &  0.98  &  1.00 \\
    4 &     7 &    2.743$-$2  &   2.751$-$2  &  0.90  &  1.00 \\
    4 &     8 &    8.326$-$2  &   8.348$-$2  &  0.90  &  1.00 \\
    4 &     9 &    3.187$-$4  &   3.139$-$4  &  0.93  &  1.02 \\
    4 &    11 &    3.356$-$2  &   3.423$-$2  &  0.94  &  0.98 \\
    4 &    17 &    7.105$-$3  &   7.107$-$3  &  0.98  &  1.00 \\
    4 &    18 &    1.064$-$1  &   1.063$-$1  &  0.98  &  1.00 \\
    4 &    19 &    5.947$-$1  &   5.945$-$1  &  0.98  &  1.00 \\
    5 &     8 &    1.638$-$4  &   1.632$-$4  &  1.50  &  1.00 \\
    5 &     9 &    1.048$-$1  &   1.061$-$1  &  1.30  &  0.99 \\
    5 &    10 &    7.063$-$2  &   7.054$-$2  &  0.58  &  1.00 \\
    5 &    12 &    1.406$-$2  &   1.456$-$2  &  0.79  &  0.97 \\
    5 &    20 &    5.462$-$1  &   5.424$-$1  &  1.00  &  1.01 \\
    6 &    15 &    1.296$-$3  &   1.365$-$3  &  1.70  &  0.95 \\
    7 &    14 &    4.245$-$4  &   4.460$-$4  &  1.70  &  0.95 \\
    7 &    15 &    2.922$-$4  &   3.098$-$4  &  1.70  &  0.94 \\
    7 &    16 &    6.034$-$4  &   6.250$-$4  &  1.60  &  0.97 \\
    8 &    13 &    1.141$-$4  &   1.106$-$4  &  0.96  &  1.03 \\
    8 &    15 &    2.565$-$4  &   2.766$-$4  &  1.80  &  0.93 \\
    8 &    16 &    1.004$-$3  &   1.048$-$3  &  1.70  &  0.96 \\
    9 &    13 &    1.181$-$2  &   1.201$-$2  &  0.62  &  0.98 \\
    9 &    15 &    6.397$-$4  &   5.647$-$4  &  0.66  &  1.13 \\
    9 &    16 &    1.847$-$6  &   2.138$-$6  &  1.90  &  0.86 \\
   10 &    13 &    2.320$-$3  &   1.955$-$3  &  0.26  &  1.19 \\
   10 &    15 &    8.195$-$5  &   5.692$-$5  &  0.15  &  1.44 \\
   11 &    14 &    3.051$-$2  &   3.080$-$2  &  1.10  &  0.99 \\
   11 &    15 &    8.808$-$2  &   8.943$-$2  &  1.10  &  0.98 \\
   11 &    16 &    1.560$-$1  &   1.573$-$1  &  1.10  &  0.99 \\
   13 &    20 &    1.154$-$1  &   1.155$-$1  &  0.79  &  1.00 \\
\hline                                                                                                        
\end{tabular}                                                                                                 
\end{table*}                                                                                                                                                                      
                                                                                                   

\clearpage
\newpage 
\setcounter{table}{3}     
\begin{table*}      
\caption{Collision strengths for resonance transitions in  Al X. ($a{\pm}b \equiv$ $a\times$10$^{{\pm}b}$).}          
\begin{tabular}{rrlllllllr}                                                                                   
\hline                                                                                                        
\hline                                                                                                        
\multicolumn{2}{c}{Transition} & \multicolumn{7}{c}{Energy (Ryd)}\\                                           
\hline                                                                                                        
  $i$ & $j$ &    50 & 100 & 150 & 200 & 250 & 300 & 350 \\                                       
\hline                                                                                                        
  1 &  2 &  1.708$-$3 &  6.687$-$4 &  3.528$-$4 &  2.180$-$4 &  1.481$-$4 &  1.070$-$4 &  8.115$-$5 \\
  1 &  3 &  5.675$-$3 &  2.584$-$3 &  1.630$-$3 &  1.195$-$3 &  9.725$-$4 &  8.336$-$4 &  7.602$-$4 \\
  1 &  4 &  8.495$-$3 &  3.323$-$3 &  1.752$-$3 &  1.082$-$3 &  7.350$-$4 &  5.314$-$4 &  4.027$-$4 \\
  1 &  5 &  1.958$+$0 &  2.275$+$0 &  2.361$+$0 &  2.286$+$0 &  2.251$+$0 &  2.206$+$0 &  2.279$+$0 \\
  1 &  6 &  4.860$-$5 &  1.852$-$5 &  1.131$-$5 &  8.698$-$6 &  7.501$-$6 &  6.860$-$6 &  6.481$-$6 \\
  1 &  7 &  1.206$-$4 &  3.403$-$5 &  1.399$-$5 &  7.060$-$6 &  4.056$-$6 &  2.540$-$6 &  1.700$-$6 \\
  1 &  8 &  2.439$-$4 &  1.062$-$4 &  7.504$-$5 &  6.460$-$5 &  6.021$-$5 &  5.810$-$5 &  5.694$-$5 \\
  1 &  9 &  1.217$-$2 &  1.273$-$2 &  1.300$-$2 &  1.315$-$2 &  1.324$-$2 &  1.331$-$2 &  1.334$-$2 \\
  1 & 10 &  4.128$-$3 &  3.532$-$3 &  3.216$-$3 &  3.020$-$3 &  2.888$-$3 &  2.793$-$3 &  2.722$-$3 \\
  1 & 11 &  1.295$-$3 &  4.340$-$4 &  2.162$-$4 &  1.297$-$4 &  8.662$-$5 &  6.202$-$5 &  4.661$-$5 \\
  1 & 12 &  6.828$-$2 &  7.410$-$2 &  7.622$-$2 &  7.737$-$2 &  7.814$-$2 &  7.872$-$2 &  7.918$-$2 \\
  1 & 13 &  7.765$-$2 &  1.371$-$1 &  1.779$-$1 &  2.084$-$1 &  2.328$-$1 &  2.532$-$1 &  2.712$-$1 \\
  1 & 14 &  3.425$-$4 &  9.731$-$5 &  4.461$-$5 &  2.538$-$5 &  1.633$-$5 &  1.135$-$5 &  8.339$-$6 \\
  1 & 15 &  5.179$-$3 &  7.740$-$3 &  9.822$-$3 &  1.144$-$2 &  1.276$-$2 &  1.386$-$2 &  1.484$-$2 \\
  1 & 16 &  1.705$-$3 &  4.843$-$4 &  2.220$-$4 &  1.263$-$4 &  8.125$-$5 &  5.649$-$5 &  4.150$-$5 \\
  1 & 17 &  1.806$-$3 &  5.195$-$4 &  2.390$-$4 &  1.360$-$4 &  8.740$-$5 &  6.076$-$5 &  4.461$-$5 \\
  1 & 18 &  3.014$-$3 &  8.719$-$4 &  4.051$-$4 &  2.338$-$4 &  1.530$-$4 &  1.088$-$4 &  8.194$-$5 \\
  1 & 19 &  4.212$-$3 &  1.211$-$3 &  5.574$-$4 &  3.172$-$4 &  2.038$-$4 &  1.417$-$4 &  1.040$-$4 \\
  1 & 20 &  1.222$-$1 &  1.606$-$1 &  1.763$-$1 &  1.846$-$1 &  1.894$-$1 &  1.926$-$1 &  1.946$-$1 \\
  1 & 21 &  7.219$-$6 &  2.127$-$6 &  9.620$-$7 &  5.386$-$7 &  3.412$-$7 &  2.344$-$7 &  1.704$-$7 \\
  1 & 22 &  5.483$-$5 &  6.518$-$5 &  7.987$-$5 &  9.248$-$5 &  1.030$-$4 &  1.121$-$4 &  1.206$-$4 \\
  1 & 23 &  3.537$-$5 &  1.043$-$5 &  4.721$-$6 &  2.644$-$6 &  1.675$-$6 &  1.151$-$6 &  8.370$-$7 \\
  1 & 24 &  2.573$-$3 &  4.370$-$3 &  5.682$-$3 &  6.696$-$3 &  7.515$-$3 &  8.208$-$3 &  8.847$-$3 \\
  1 & 25 &  9.645$-$5 &  6.424$-$5 &  4.745$-$5 &  3.707$-$5 &  3.011$-$5 &  2.515$-$5 &  2.146$-$5 \\
  1 & 26 &  7.385$-$5 &  2.769$-$5 &  1.486$-$5 &  9.490$-$6 &  6.706$-$6 &  5.058$-$6 &  3.993$-$6 \\
  1 & 27 &  1.200$-$4 &  4.268$-$5 &  2.257$-$5 &  1.472$-$5 &  1.093$-$5 &  8.830$-$6 &  7.552$-$6 \\
  1 & 28 &  1.582$-$4 &  5.301$-$5 &  2.551$-$5 &  1.476$-$5 &  9.546$-$6 &  6.648$-$6 &  4.885$-$6 \\
  1 & 29 &  5.365$-$5 &  1.732$-$5 &  8.380$-$6 &  4.941$-$6 &  3.270$-$6 &  2.334$-$6 &  1.754$-$6 \\
  1 & 30 &  9.022$-$6 &  3.664$-$6 &  2.649$-$6 &  2.333$-$6 &  2.205$-$6 &  2.145$-$6 &  2.113$-$6 \\
  1 & 31 &  2.166$-$5 &  5.233$-$6 &  2.040$-$6 &  1.014$-$6 &  5.849$-$7 &  3.716$-$7 &  2.533$-$7 \\
  1 & 32 &  4.262$-$5 &  1.928$-$5 &  1.521$-$5 &  1.402$-$5 &  1.356$-$5 &  1.334$-$5 &  1.323$-$5 \\
  1 & 33 &  9.028$-$5 &  4.142$-$5 &  3.065$-$5 &  2.549$-$5 &  2.216$-$5 &  1.973$-$5 &  1.784$-$5 \\
  1 & 34 &  1.035$-$4 &  2.745$-$5 &  1.348$-$5 &  8.746$-$6 &  6.617$-$6 &  5.488$-$6 &  4.823$-$6 \\
  1 & 35 &  1.284$-$4 &  9.889$-$5 &  8.740$-$5 &  7.823$-$5 &  7.063$-$5 &  6.430$-$5 &  5.898$-$5 \\
  1 & 36 &  1.281$-$4 &  3.151$-$5 &  1.366$-$5 &  7.577$-$6 &  4.813$-$6 &  3.330$-$6 &  2.443$-$6 \\
  1 & 37 &  1.036$-$3 &  1.273$-$3 &  1.348$-$3 &  1.377$-$3 &  1.389$-$3 &  1.394$-$3 &  1.396$-$3 \\
  1 & 38 &  5.883$-$5 &  6.940$-$5 &  8.107$-$5 &  9.052$-$5 &  9.820$-$5 &  1.047$-$4 &  1.108$-$4 \\
  1 & 39 &  2.497$-$5 &  5.629$-$6 &  2.388$-$6 &  1.365$-$6 &  9.207$-$7 &  6.864$-$7 &  5.460$-$7 \\
  1 & 40 &  2.571$-$5 &  6.975$-$6 &  4.314$-$6 &  3.632$-$6 &  3.403$-$6 &  3.318$-$6 &  3.288$-$6 \\
  1 & 41 &  1.322$-$4 &  3.616$-$5 &  1.639$-$5 &  9.316$-$6 &  6.021$-$6 &  4.225$-$6 &  3.142$-$6 \\
  1 & 42 &  9.320$-$5 &  3.969$-$5 &  3.102$-$5 &  2.927$-$5 &  2.931$-$5 &  2.995$-$5 &  3.090$-$5 \\
  1 & 43 &  2.743$-$5 &  7.464$-$6 &  3.351$-$6 &  1.881$-$6 &  1.199$-$6 &  8.287$-$7 &  6.066$-$7 \\
  1 & 44 &  5.905$-$4 &  5.783$-$4 &  5.762$-$4 &  5.761$-$4 &  5.766$-$4 &  5.774$-$4 &  5.783$-$4 \\
  1 & 45 &  1.108$-$3 &  1.209$-$3 &  1.252$-$3 &  1.280$-$3 &  1.300$-$3 &  1.316$-$3 &  1.330$-$3 \\
  1 & 46 &  6.973$-$3 &  1.000$-$2 &  1.198$-$2 &  1.347$-$2 &  1.466$-$2 &  1.566$-$2 &  1.658$-$2 \\
  1 & 47 &  4.959$-$4 &  1.521$-$4 &  7.309$-$5 &  4.301$-$5 &  2.842$-$5 &  2.010$-$5 &  1.508$-$5 \\
  1 & 48 &  1.337$-$2 &  1.490$-$2 &  1.547$-$2 &  1.578$-$2 &  1.598$-$2 &  1.613$-$2 &  1.625$-$2 \\
  1 & 49 &  1.600$-$4 &  4.049$-$5 &  1.768$-$5 &  9.791$-$6 &  6.201$-$6 &  4.252$-$6 &  3.104$-$6 \\
  1 & 50 &  6.261$-$4 &  3.797$-$4 &  3.866$-$4 &  4.202$-$4 &  4.553$-$4 &  4.886$-$4 &  5.208$-$4 \\
\hline                                                                                                        
\end{tabular}                                                                                                 
\end{table*}                                                                                                  
\newpage                                                                                                      
\clearpage                                                                                                    
\setcounter{table}{3}                                                                                         
\begin{table*}                                                                                                
\caption{Collision strengths for resonance transitions in  Al X. ($a{\pm}b \equiv$ $a\times$10$^{{\pm}b}$).}            
\begin{tabular}{rrlllllllr}                                                                                  
\hline                                                                                                        
\hline                                                                                                        
\multicolumn{2}{c}{Transition} & \multicolumn{7}{c}{Energy (Ryd)} \\                                           
\hline                                                                                                        
  $i$ & $j$ &    50 & 100 & 150 & 200 & 250 & 300 & 350 \\                                             
\hline                                                                                                        
  1 & 51 &  7.968$-$4 &  2.016$-$4 &  8.798$-$5 &  4.874$-$5 &  3.086$-$5 &  2.116$-$5 &  1.545$-$5 \\
  1 & 52 &  1.720$-$2 &  2.989$-$2 &  3.864$-$2 &  4.533$-$2 &  5.071$-$2 &  5.529$-$2 &  5.950$-$2 \\
  1 & 53 &  7.155$-$4 &  1.973$-$4 &  8.952$-$5 &  5.061$-$5 &  3.239$-$5 &  2.245$-$5 &  1.645$-$5 \\
  1 & 54 &  1.193$-$3 &  3.305$-$4 &  1.511$-$4 &  8.643$-$5 &  5.615$-$5 &  3.963$-$5 &  2.967$-$5 \\
  1 & 55 &  1.667$-$3 &  4.597$-$4 &  2.086$-$4 &  1.179$-$4 &  7.546$-$5 &  5.229$-$5 &  3.833$-$5 \\
  1 & 56 &  2.259$-$2 &  3.019$-$2 &  3.328$-$2 &  3.491$-$2 &  3.590$-$2 &  3.657$-$2 &  3.702$-$2 \\
  1 & 57 &  3.191$-$4 &  6.620$-$5 &  2.659$-$5 &  1.410$-$5 &  8.701$-$6 &  5.896$-$6 &  4.257$-$6 \\
  1 & 58 &  4.472$-$4 &  9.363$-$5 &  3.828$-$5 &  2.085$-$5 &  1.330$-$5 &  9.387$-$6 &  7.101$-$6 \\
  1 & 59 &  5.734$-$4 &  1.189$-$4 &  4.774$-$5 &  2.533$-$5 &  1.562$-$5 &  1.059$-$5 &  7.643$-$6 \\
  1 & 60 &  7.039$-$3 &  8.601$-$3 &  9.011$-$3 &  9.177$-$3 &  9.265$-$3 &  9.323$-$3 &  9.370$-$3 \\
  1 & 61 &  3.291$-$6 &  9.160$-$7 &  4.092$-$7 &  2.283$-$7 &  1.447$-$7 &  9.946$-$8 &  7.254$-$8 \\
  1 & 62 &  2.009$-$5 &  1.559$-$5 &  1.624$-$5 &  1.743$-$5 &  1.860$-$5 &  1.965$-$5 &  2.064$-$5 \\
  1 & 63 &  1.612$-$5 &  4.526$-$6 &  2.029$-$6 &  1.134$-$6 &  7.198$-$7 &  4.955$-$7 &  3.618$-$7 \\
  1 & 64 &  1.437$-$4 &  1.640$-$4 &  1.868$-$4 &  2.062$-$4 &  2.225$-$4 &  2.363$-$4 &  2.487$-$4 \\
  1 & 65 &  2.305$-$5 &  1.064$-$5 &  7.057$-$6 &  5.311$-$6 &  4.259$-$6 &  3.551$-$6 &  3.039$-$6 \\
  1 & 66 &  2.114$-$5 &  8.895$-$6 &  5.593$-$6 &  4.080$-$6 &  3.209$-$6 &  2.640$-$6 &  2.238$-$6 \\
  1 & 67 &  3.157$-$5 &  1.138$-$5 &  6.698$-$6 &  4.922$-$6 &  4.062$-$6 &  3.573$-$6 &  3.271$-$6 \\
  1 & 68 &  4.064$-$5 &  1.227$-$5 &  5.681$-$6 &  3.215$-$6 &  2.051$-$6 &  1.414$-$6 &  1.032$-$6 \\
  1 & 69 &  2.131$-$5 &  6.721$-$6 &  3.437$-$6 &  2.170$-$6 &  1.538$-$6 &  1.171$-$6 &  9.333$-$7 \\
  1 & 70 &  6.768$-$6 &  4.381$-$6 &  4.084$-$6 &  4.052$-$6 &  4.074$-$6 &  4.108$-$6 &  4.141$-$6 \\
  1 & 71 &  1.644$-$5 &  4.418$-$6 &  1.942$-$6 &  1.072$-$6 &  6.746$-$7 &  4.634$-$7 &  3.363$-$7 \\
  1 & 72 &  1.916$-$5 &  6.966$-$6 &  4.749$-$6 &  4.005$-$6 &  3.657$-$6 &  3.454$-$6 &  3.324$-$6 \\
  1 & 73 &  3.402$-$5 &  1.220$-$5 &  7.824$-$6 &  6.030$-$6 &  5.020$-$6 &  4.350$-$6 &  3.863$-$6 \\
  1 & 74 &  4.131$-$5 &  1.177$-$5 &  6.655$-$6 &  4.987$-$6 &  4.262$-$6 &  3.892$-$6 &  3.683$-$6 \\
  1 & 75 &  4.041$-$5 &  1.988$-$5 &  1.492$-$5 &  1.245$-$5 &  1.084$-$5 &  9.658$-$6 &  8.739$-$6 \\
  1 & 76 &  1.185$-$4 &  1.234$-$4 &  1.223$-$4 &  1.196$-$4 &  1.167$-$4 &  1.141$-$4 &  1.117$-$4 \\
  1 & 77 &  5.048$-$5 &  1.179$-$5 &  5.015$-$6 &  2.752$-$6 &  1.737$-$6 &  1.196$-$6 &  8.749$-$7 \\
  1 & 78 &  6.758$-$5 &  8.629$-$5 &  1.020$-$4 &  1.144$-$4 &  1.247$-$4 &  1.335$-$4 &  1.412$-$4 \\
  1 & 79 &  2.511$-$5 &  7.313$-$6 &  4.018$-$6 &  2.817$-$6 &  2.213$-$6 &  1.846$-$6 &  1.598$-$6 \\
  1 & 80 &  1.226$-$5 &  7.371$-$6 &  5.516$-$6 &  4.402$-$6 &  3.653$-$6 &  3.117$-$6 &  2.717$-$6 \\
  1 & 81 &  3.206$-$6 &  6.064$-$7 &  2.584$-$7 &  1.492$-$7 &  1.003$-$7 &  7.370$-$8 &  5.740$-$8 \\
  1 & 82 &  1.084$-$5 &  1.346$-$5 &  1.516$-$5 &  1.622$-$5 &  1.691$-$5 &  1.738$-$5 &  1.773$-$5 \\
  1 & 83 &  9.779$-$6 &  7.992$-$6 &  8.248$-$6 &  8.531$-$6 &  8.752$-$6 &  8.923$-$6 &  9.058$-$6 \\
  1 & 84 &  1.521$-$5 &  3.767$-$6 &  2.166$-$6 &  1.740$-$6 &  1.588$-$6 &  1.525$-$6 &  1.498$-$6 \\
  1 & 85 &  4.825$-$5 &  1.223$-$5 &  5.396$-$6 &  3.040$-$6 &  1.964$-$6 &  1.385$-$6 &  1.038$-$6 \\
  1 & 86 &  3.925$-$5 &  1.623$-$5 &  1.289$-$5 &  1.236$-$5 &  1.253$-$5 &  1.291$-$5 &  1.336$-$5 \\
  1 & 87 &  1.247$-$5 &  3.170$-$6 &  1.382$-$6 &  7.625$-$7 &  4.802$-$7 &  3.291$-$7 &  2.394$-$7 \\
  1 & 88 &  7.156$-$6 &  3.137$-$6 &  2.168$-$6 &  1.672$-$6 &  1.362$-$6 &  1.148$-$6 &  9.925$-$7 \\
  1 & 89 &  8.816$-$6 &  5.361$-$6 &  5.167$-$6 &  5.218$-$6 &  5.295$-$6 &  5.367$-$6 &  5.429$-$6 \\
  1 & 90 &  9.487$-$6 &  1.901$-$6 &  8.236$-$7 &  4.637$-$7 &  2.980$-$7 &  2.076$-$7 &  1.528$-$7 \\
  1 & 91 &  8.631$-$6 &  2.003$-$6 &  9.824$-$7 &  6.338$-$7 &  4.655$-$7 &  3.675$-$7 &  3.037$-$7 \\
  1 & 92 &  2.831$-$5 &  3.422$-$5 &  3.858$-$5 &  4.132$-$5 &  4.309$-$5 &  4.429$-$5 &  4.516$-$5 \\
  1 & 93 &  4.208$-$5 &  4.565$-$5 &  4.894$-$5 &  5.115$-$5 &  5.270$-$5 &  5.384$-$5 &  5.472$-$5 \\
  1 & 94 &  3.797$-$6 &  6.927$-$7 &  2.597$-$7 &  1.317$-$7 &  7.876$-$8 &  5.216$-$8 &  3.701$-$8 \\
  1 & 95 &  2.085$-$4 &  2.169$-$4 &  2.255$-$4 &  2.315$-$4 &  2.359$-$4 &  2.394$-$4 &  2.421$-$4 \\
  1 & 96 &  9.353$-$5 &  1.296$-$4 &  1.483$-$4 &  1.594$-$4 &  1.665$-$4 &  1.712$-$4 &  1.746$-$4 \\
  1 & 97 &  1.887$-$4 &  1.828$-$4 &  1.831$-$4 &  1.849$-$4 &  1.868$-$4 &  1.887$-$4 &  1.905$-$4 \\
  1 & 98 &  1.837$-$3 &  2.641$-$3 &  3.166$-$3 &  3.567$-$3 &  3.894$-$3 &  4.171$-$3 &  4.414$-$3 \\
\hline                                                                                                        
\end{tabular}                                                                                                 
\end{table*}                                                                                                                                                                                                 
                   
 
 \newpage                                                                                                      
\clearpage    
\setcounter{table}{5}                                                                                         
\begin{table*}                                                                                                
\caption{Comparisons of effective collision strengths for transitions among the lowest 10 levels of Al X. $a{\pm}b \equiv a{\times}$10$^{{\pm}b}$.}

\begin{tabular}{rrlllllllll} \hline  \hline
\multicolumn{2}{c}{Transition} & \multicolumn{3}{c}{RM (log T$_e$, K)} & \multicolumn{3}{c}{DARC (log T$_e$, K)} \\
 \hline
   I  &      J    &      5.90      &     6.10       &     6.30    &   5.90        &     6.10      &    6.30   \\
  \hline
   1  &      2    &   7.824$-$3    &   6.631$-$3    &	5.524$-$3 &  1.161$-$2  &  1.076$-$2  &  9.407$-$3    \\
   1  &      3    &   2.347$-$2    &   1.989$-$2    &	1.657$-$2 &  3.528$-$2  &  3.275$-$2  &  2.870$-$2    \\
   1  &      4    &   3.912$-$2    &   3.316$-$2    &	2.762$-$2 &  5.881$-$2  &  5.444$-$2  &  4.753$-$2    \\
   1  &      5    &   1.296$-$0    &   1.368$-$0    &	1.459$-$0 &  1.280$-$0  &  1.346$-$0  &  1.433$-$0    \\
   1  &      6    &   2.038$-$4    &   1.807$-$4    &	1.552$-$4 &  6.151$-$4  &  6.846$-$4  &  6.534$-$4    \\
   1  &      7    &   6.113$-$4    &   5.420$-$4    &	4.657$-$4 &  1.744$-$3  &  1.930$-$3  &  1.834$-$3    \\
   1  &      8    &   1.019$-$3    &   9.033$-$4    &	7.761$-$4 &  2.827$-$3  &  3.075$-$3  &  2.902$-$3    \\
   1  &      9    &   1.281$-$2    &   1.249$-$2    &	1.220$-$2 &  1.711$-$2  &  1.736$-$2  &  1.687$-$2    \\
   1  &     10    &   4.581$-$3    &   4.495$-$3    &	4.383$-$3 &  7.199$-$3  &  7.312$-$3  &  7.000$-$3    \\
   2  &      3    &   1.467$-$1    &   1.142$-$1    &	8.695$-$2 &  1.504$-$1  &  1.258$-$1  &  1.023$-$1    \\
   2  &      4    &   1.316$-$1    &   1.009$-$1    &	7.795$-$2 &  1.139$-$1  &  9.819$-$2  &  8.375$-$2    \\
   2  &      5    &   1.749$-$2    &   1.472$-$2    &	1.231$-$2 &  3.116$-$2  &  2.919$-$2  &  2.533$-$2    \\
   2  &      6    &   4.106$-$3    &   3.683$-$3    &	3.211$-$3 &  5.641$-$3  &  5.463$-$3  &  4.910$-$3    \\
   2  &      7    &   6.600$-$1    &   6.922$-$1    &	7.328$-$1 &  6.409$-$1  &  6.727$-$1  &  7.153$-$1    \\
   2  &      8    &   5.527$-$3    &   4.973$-$3    &	4.356$-$3 &  1.148$-$2  &  1.163$-$2  &  1.061$-$2    \\
   2  &      9    &   9.489$-$3    &   8.527$-$3    &	7.446$-$3 &  1.695$-$2  &  1.700$-$2  &  1.534$-$2    \\
   2  &     10    &   1.139$-$3    &   1.034$-$3    &	9.112$-$4 &  2.704$-$3  &  2.638$-$3  &  2.307$-$3    \\
   3  &      4    &   4.821$-$1    &   3.721$-$1    &	2.844$-$1 &  5.072$-$1  &  4.196$-$1  &  3.430$-$1    \\
   3  &      5    &   5.247$-$2    &   4.417$-$2    &	3.693$-$2 &  9.302$-$2  &  8.719$-$2  &  7.566$-$2    \\
   3  &      6    &   6.561$-$1    &   6.892$-$1    &	7.307$-$1 &  6.438$-$1  &  6.755$-$1  &  7.177$-$1    \\
   3  &      7    &   5.150$-$1    &   5.408$-$1    &	5.734$-$1 &  5.057$-$1  &  5.292$-$1  &  5.588$-$1    \\
   3  &      8    &   8.488$-$1    &   8.973$-$1    &	9.582$-$1 &  8.240$-$1  &  8.643$-$1  &  9.153$-$1    \\
   3  &      9    &   2.847$-$2    &   2.558$-$2    &	2.234$-$2 &  5.341$-$2  &  5.340$-$2  &  4.812$-$2    \\
   3  &     10    &   3.417$-$3    &   3.102$-$3    &	2.734$-$3 &  8.667$-$3  &  8.480$-$3  &  7.434$-$3    \\
   4  &      5    &   8.744$-$2    &   7.361$-$2    &	6.156$-$2 &  1.553$-$1  &  1.452$-$1  &  1.258$-$1    \\
   4  &      6    &   5.508$-$3    &   4.959$-$3    &	4.347$-$3 &  9.998$-$3  &  1.008$-$2  &  9.185$-$3    \\
   4  &      7    &   8.499$-$1    &   8.984$-$1    &	9.590$-$1 &  8.273$-$1  &  8.675$-$1  &  9.186$-$1    \\
   4  &      8    &   2.517$-$0    &   2.642$-$0    &	2.799$-$0 &  2.437$-$0  &  2.556$-$0  &  2.712$-$0    \\
   4  &      9    &   4.744$-$2    &   4.263$-$2    &	3.723$-$2 &  1.039$-$1  &  1.037$-$1  &  9.444$-$2    \\
   4  &     10    &   5.694$-$3    &   5.170$-$3    &	4.556$-$3 &  1.587$-$2  &  1.547$-$2  &  1.351$-$2    \\
   5  &      6    &   7.206$-$3    &   6.169$-$3    &	5.188$-$3 &  1.264$-$2  &  1.193$-$2  &  1.057$-$2    \\
   5  &      7    &   2.162$-$2    &   1.851$-$2    &	1.556$-$2 &  3.682$-$2  &  3.439$-$2  &  3.002$-$2    \\
   5  &      8    &   3.603$-$2    &   3.084$-$2    &	2.594$-$2 &  7.226$-$2  &  6.812$-$2  &  6.101$-$2    \\
   5  &      9    &   3.629$-$0    &   3.816$-$0    &	4.049$-$0 &  3.517$-$0  &  3.686$-$0  &  3.913$-$0    \\
   5  &     10    &   1.161$-$0    &   1.225$-$0    &	1.305$-$0 &  1.143$-$0  &  1.204$-$0  &  1.284$-$0    \\
   6  &      7    &   7.308$-$2    &   6.478$-$2    &	5.610$-$2 &  1.016$-$1  &  9.890$-$2  &  8.872$-$2    \\
   6  &      8    &   4.909$-$2    &   4.674$-$2    &	4.474$-$2 &  6.447$-$2  &  6.499$-$2  &  6.194$-$2    \\
   6  &      9    &   3.784$-$2    &   3.390$-$2    &	2.950$-$2 &  4.965$-$2  &  4.779$-$2  &  4.244$-$2    \\
   6  &     10    &   4.239$-$3    &   3.724$-$3    &	3.158$-$3 &  1.008$-$2  &  9.699$-$3  &  8.374$-$3    \\
   7  &      8    &   2.009$-$1    &   1.865$-$1    &	1.731$-$1 &  2.703$-$1  &  2.677$-$1  &  2.480$-$1    \\
   7  &      9    &   1.135$-$1    &   1.017$-$1    &	8.850$-$2 &  1.553$-$1  &  1.498$-$1  &  1.332$-$1    \\
   7  &     10    &   1.272$-$2    &   1.117$-$2    &	9.473$-$3 &  3.065$-$2  &  2.926$-$2  &  2.514$-$2    \\
   8  &      9    &   1.892$-$1    &   1.695$-$1    &	1.475$-$1 &  2.696$-$1  &  2.592$-$1  &  2.306$-$1    \\
   8  &     10    &   2.119$-$2    &   1.862$-$2    &	1.579$-$2 &  5.313$-$2  &  5.059$-$2  &  4.336$-$2    \\
   9  &     10    &   9.438$-$2    &   9.645$-$2    &	9.929$-$2 &  1.141$-$1  &  1.172$-$1  &  1.181$-$1    \\
 \hline
\end{tabular} 

\begin{flushleft}
{\small
RM: Earlier interpolated results of  \cite{fpk1} \\ 
DARC: Present results from the DARC code  \\
}
\end{flushleft}
\end{table*}


\begin{thebibliography}{99}
\bibitem[\protect\citeauthoryear{Aggarwal \& Keenan}{2004}]{mo34}
Aggarwal K. M.,   Keenan F. P.,  2004, Phys. Scr., 69, 176 
\bibitem[\protect\citeauthoryear{Aggarwal \& Keenan}{2008}]{ni11}
Aggarwal K. M.,  Keenan F. P., 2008,  Eur. Phys. J.,  D 46,  205 
\bibitem[\protect\citeauthoryear{Aggarwal \& Keenan}{2012a}]{mgxi}
Aggarwal K. M.,   Keenan F. P., 2012a,  Phys. Scr., 85, 025305 
\bibitem[\protect\citeauthoryear{Aggarwal \& Keenan}{2012b}]{caxix}
Aggarwal K. M.,   Keenan F. P., 2012b, Phys. Scr.,  85, 025306 
\bibitem[\protect\citeauthoryear{Aggarwal \& Keenan}{2012c}]{tixix}
Aggarwal K. M.,    Keenan F. P., 2012c, Phys. Scr.,  86,  055301 
\bibitem[\protect\citeauthoryear{Aggarwal \& Keenan}{2013}]{fe25}
Aggarwal K. M.,   Keenan F. P., 2013, Phys. Scr., 87, 055302 
\bibitem[\protect\citeauthoryear{Aggarwal, Keenan \& Lawson}{Aggarwal et al.}{2008}]{kr35}
Aggarwal K. M.,   Keenan F. P., Lawson K. D., 2008, At. Data Nucl. Data Tables, 94, 323 
 \bibitem[\protect\citeauthoryear{Aggarwal, Keenan \& Lawson}{Aggarwal et al.}{2010}]{xe54}
Aggarwal K. M.,   Keenan F. P., Lawson K. D., 2010, At. Data Nucl. Data Tables, 96, 123 
 \bibitem[\protect\citeauthoryear{Aggarwal et al.}{2007}]{fe15}
Aggarwal  K. M., Tayal V., Gupta G. P.,   Keenan F. P., 2007,  At. Data Nucl. Data Tables, 93, 615  
\bibitem[\protect\citeauthoryear{Andersson et al.}{2009}]{ma}
Andersson M., Zou Y.,  Hutton R.,   Brage T., 2009, Phys. Rev. A 79, 032501  
\bibitem[\protect\citeauthoryear{Bar-Shalom, Klapisch \& Oreg}{Bar-Shalom et al.}{2001}]{hullac}
Bar-Shalom A., Klapisch M.  Oreg, J., 2001, J. Quant. Spect. Rad. Trans., 71, 169  
\bibitem[\protect\citeauthoryear{Bryans, Landi \& Savin}{Bryans et al.}{2009}]{pb}
Bryans P., Landi E.,    Savin D. W., 2009,  ApJ,  691, 1540 
\bibitem[\protect\citeauthoryear{Burgess \& Sheorey}{1974}]{ab}
Burgess A.,   Sheorey V. B.,  1974, J. Phys.,  B7,  2403 
\bibitem[\protect\citeauthoryear{Burgess \& Tully}{1992}]{bt}
Burgess A.,    Tully J. A., 1992, A\&A,   254, 436 
\bibitem[\protect\citeauthoryear{Garstang}{1968}]{rhg}
Garstang R. H., 1968,	J. Phys. , B1,  847 
\bibitem[\protect\citeauthoryear{Grant et al.}{1980}]{grasp0}
Grant I. P., McKenzie B. J.,  Norrington P. H.,  Mayers D. F.,   Pyper N. C., 1980,	Comput. Phys. Commun.,  21,  207  
\bibitem[\protect\citeauthoryear{Gu}{2008}]{fac}
Gu M. F., 2008, Can. J. Phys., 86, 675 
\bibitem[\protect\citeauthoryear{Gu, Beiersdorfer \& Lepson}{Gu et al.}{2011}]{gu1}
Gu M. F., Beiersdorfer P.,  Lepson J. K., 2011, ApJ, 732, 91 
\bibitem[\protect\citeauthoryear{Keenan et al.}{1986}]{fpk1}
Keenan F. P., Berrington K. A., Burke P. G., Dufton P. L.,  Kingston A. E., 1986, Phys. Scr., 34, 216 
\bibitem[\protect\citeauthoryear{Landi et al.}{2001}]{el1}
Landi E., Doron R., Feldman U., Doscheck G. A., 2001, ApJ, 556, 912 
\bibitem[\protect\citeauthoryear{Martin \& Zalubas}{1979}]{mz}
Martin   W.  C.,   Zalubas R., 1979,   J. Phys. Chem. Ref. Data, 8, 817 
\bibitem[\protect\citeauthoryear{Tr{\" a}bert \& Heckmann}{1980}]{et1}
Tr{\" a}bert E.,  Heckmann P. H., 1980,  Phys. Scr., 22, 489 
\bibitem[\protect\citeauthoryear{Wilhelm et al.}{1995}]{wk1}
Wilhelm K. et al., 1995, Sol. Phys., 162, 189 
\bibitem[\protect\citeauthoryear{Woodgate}{1970}]{wood}
Woodgate G. K., 1970 {\em Elementary Atomic Structure}, New York: McGraw-Hill 
\bibitem[\protect\citeauthoryear{Zhang \& Sampson}{1992}]{zs92}
Zhang H.,   Sampson D. H., 1992,  At. Data Nucl. Data Tables, 52, 143 

\end{thebibliography}
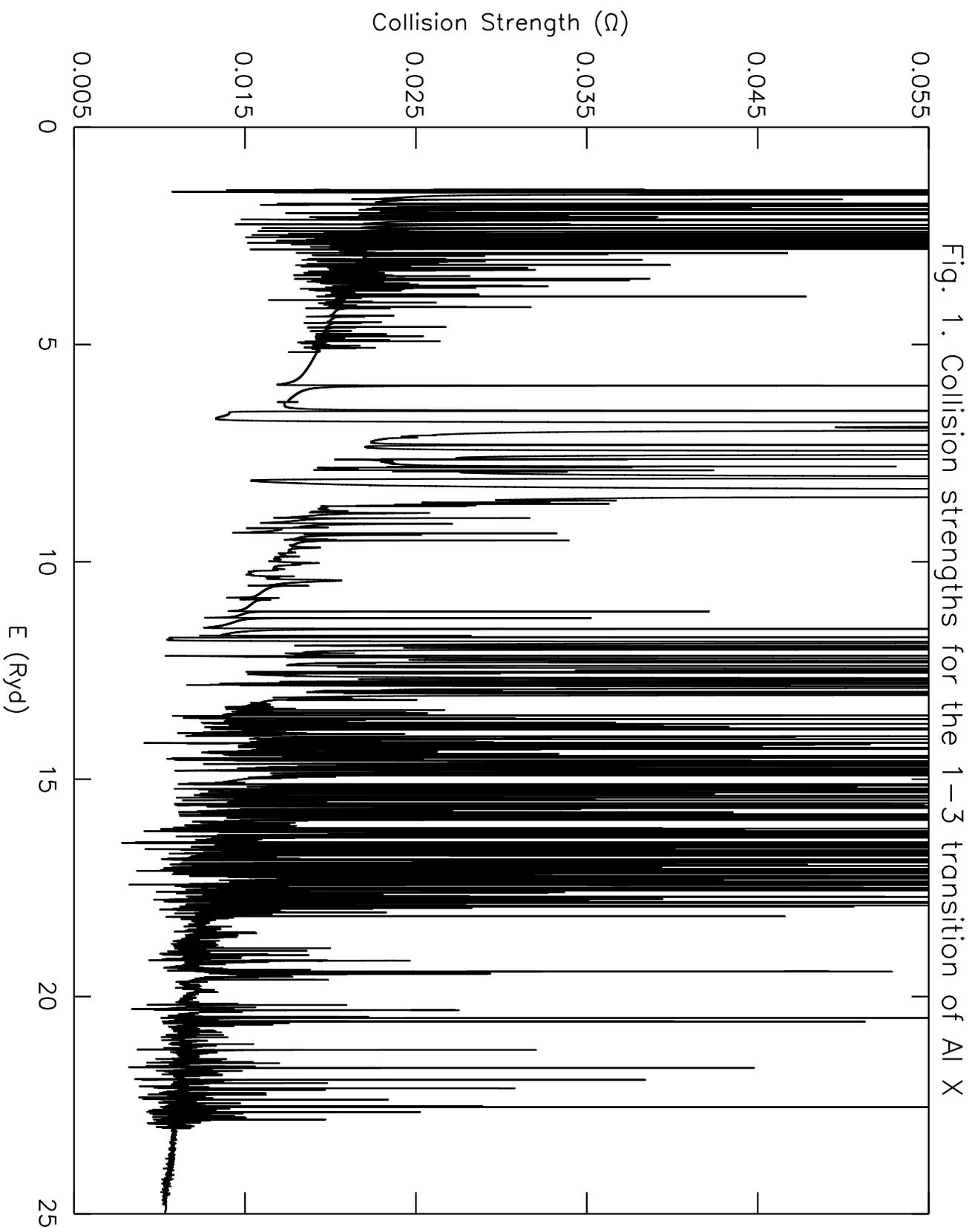
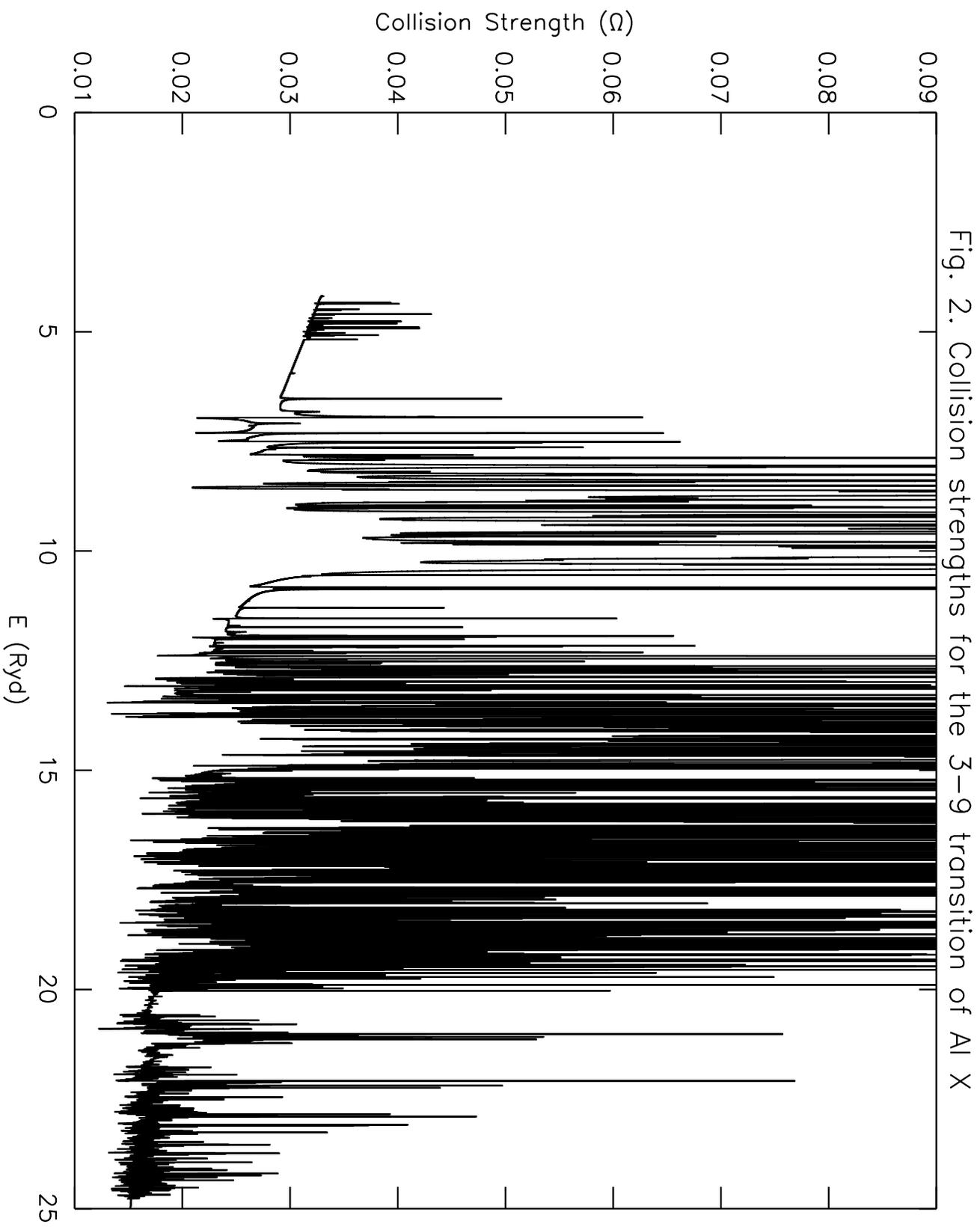
\end{document}